\documentclass[12pt]{article}
\usepackage{bm}
\usepackage{graphicx}

\title{QCD AGAINST BLACK HOLES OF A STAR MASS ?}
 
\vspace{5mm}

\author{Ilya I. Royzen}
\date{}
\begin{document}
\maketitle \centerline{\em{P.N. Lebedev Physical Institute of RAS}},

\centerline{e-mail: $<royzen@lpi.ru>$}
\addtolength{\baselineskip}{6pt}
\vspace*{20mm}
\begin{abstract}



 Along with compacting baryon (neutron) spacing in a neutron star (NS), two very 
important factors come into play side by side: the lack of the NS gravitational 
self-stabilization against shutting to black hole (BH) and the phase transition 
- color deconfinement and QCD-vacuum reconstruction - within the nuclear matter 
the NS is composed of. That is why both phenomena should be taken into          
account at once, as the gravitational collapse is considered. Since, under the 
above transition, the hadronic-phase (HPh) vacuum (filled up with gluon- and 
chiral $q\bar q$-condensates) turns into the ''empty''
(perturbation) subhadronic-phase (SHPh) one and, thus, the formerly (very high)  
pressure falls down rather abruptly, the formerly cold nuclear medium starts 
imploding almost freely into the new vacuum. If the star mass is sufficiently 
large, then this implosion is shown to result in an enormous heating - up to 
the temperature about 100 MeV or, may be, even higher - and growth of the inner 
pressure due to degeneracy breaking and multiple $q\bar q$-pair production 
which withstands the gravitational compression (remind that the 
highest temperatures of supernovae bursts, as well as of the ''normal'' NS, are,
at least, of one order lower). As a consequence, a ''flaming wall'' is, most 
probably, emerged on the way of further collapsing which prevents the NS to 
evolve towards the BH horizon appearance. At the same time, it could give rise 
to the most powerful GRBs produced by some very distant (young) stars.

\end{abstract}




\newpage
\section{Two incompatible mechanisms of neutron star instability}

Two mechanisms underlying a compact star instability are confronted below
which make the star to evolve in absolutely alternative ways: the 
first one implies HPh $\to$ SHPh transition within nuclear matter (it is 
described here in more detail) and the second one, which is rather familiar, is 
NS shutting to BH. Thus, the  main point is to understand, which one is 
activated before.

\subsection{Phase transition in nuclear medium}

Schematically, this transition is depicted as follows: \\

\qquad\qquad\qquad \,\, {QCD HPh  \qquad\qquad\qquad  
{$\Longleftrightarrow$} \qquad\qquad\quad QCD SHPh}\\ 

\hspace*{3.5cm} {$\Downarrow$ \hspace*{7cm}  $\Downarrow$ }\\

\hspace*{1.0cm} {$P^0_{vac}=-\varepsilon^0_{vac}\simeq
5\,10^{-3}$ {GeV$^4$} \quad\qquad $\Longleftrightarrow$ \quad\qquad $P_{vac}=
-\varepsilon_{vac}\rightarrow\,0$}\\

\hspace*{3.5cm} {$\Downarrow$ \hspace*{7cm}  $\Downarrow$ }\\

\hspace*{0.0cm}$P_{tot}\simeq\,P^0_{vac}\,\,
\lbrack rarefied\,\,gas\,\,of\,\,nucleons\rbrack$ \quad\quad 
$\Longleftrightarrow$ \quad\quad {$P_{tot}=P_{vac}\,\,+\,\,P$} \\

\hspace*{0.0cm}$\varepsilon_{tot}\simeq\,\varepsilon^0_{vac}\,\,
\lbrack rarefied\,\,gas\,\,of\,\,nucleons\rbrack$ \quad\quad 
$\Longleftrightarrow$ \quad\quad {$\varepsilon_{tot}=\varepsilon_{vac}\,\,+\,
\varepsilon$} \\

\noindent Here ($\varepsilon^0_{vac}$, $P^0_{vac}$) and 
($\varepsilon_{vac}$, $P_{vac}$)
stand for the vacuum parameters (energy density, pressure) in HPh and SHPh, 
respectively,  
while ($\varepsilon,\,\,P$) are the particle ones, and 
($\varepsilon_{tot},\,\,P_{tot}$) are the overall energy density and 
pressure within the nuclear medium. It is worth emphasizing that 
$|\varepsilon^0_{vac}|\,\simeq\,\varepsilon_{n}$, the latter being the particle 
energy density of
''close packed'' nucleons (neutrons) which is somewhat - most probably,
(25-30)\% - higher than the mean intrinsic energy (mass) density within a free 
nucleon itself.
  
Below, we consider two conceivable scenarios of this phase transition 
\cite{R_2008,R_2009,R-F-Ch} - the hard scenario, when the HPh transforms at 
some fixed density (pressure) directly 
(stepwise) into the current quark state (this is the ''conventional'' phase 
transition), and the soft one, which admits an intermediate state in between
(a kind of crossover).
The latter asks for the notion of deconfined 
dynamical quarks (valons) - quasi-particles of non-fixed mass, which 
diminishes along with the density (pressure) increase.
It is shown below that both scenarios result in developing strong 
instability under the phase transformation.\\ \\ 

{\bf\boldmath 1.\quad$Hard\,\,scenario$:

\hspace*{1cm}  stepwise transition to the ''empty''  
vacuum where $P_{vac}\,=\,\varepsilon_{vac}\,\equiv\,0$}\\ \\
This scenario implies that the current quarks 
(almost massless ($u,d$)- \,and \,$\sim$150-MeV $s$-quark) are emerged 
promptly as neutrons crush down. It is illustrated in Fig.1 \cite{R_2008}
where the HPh vacuum condensate pressure $P^0_{vac}$ is confronted with that of the 
degenerate current quark gas at the different particle number densities,  
which is represented by the quark specific volume $\langle v\rangle$. It is
obvious, that, 
in the framework of this scenario, the transition into {\it degenerate 
(''cold'') quark gas} is ruled out - in fact, the matter starts collapsing into 
the new zero-rigidity (''empty'') vacuum, what gives rise to an enormous 
heating (see an 
estimate below) of the nuclear medium just after the phase transition point is 
passed through \footnote{Two pressures - the HPh-vacuum pressure and the 
pressure of degenerate perfect SHPh-particle gas - would equate only at the
quark number density (point {\it B} in Fig.1), which is 3-4 times as high as 
the phase transition point one. It is worth noting that the neutrinos get 
essentially stuck under relevant densities and, thus, there is no way for an 
''instant'' energy release from the star interior.}.  

\begin{figure}[h]
\includegraphics[width=10cm]{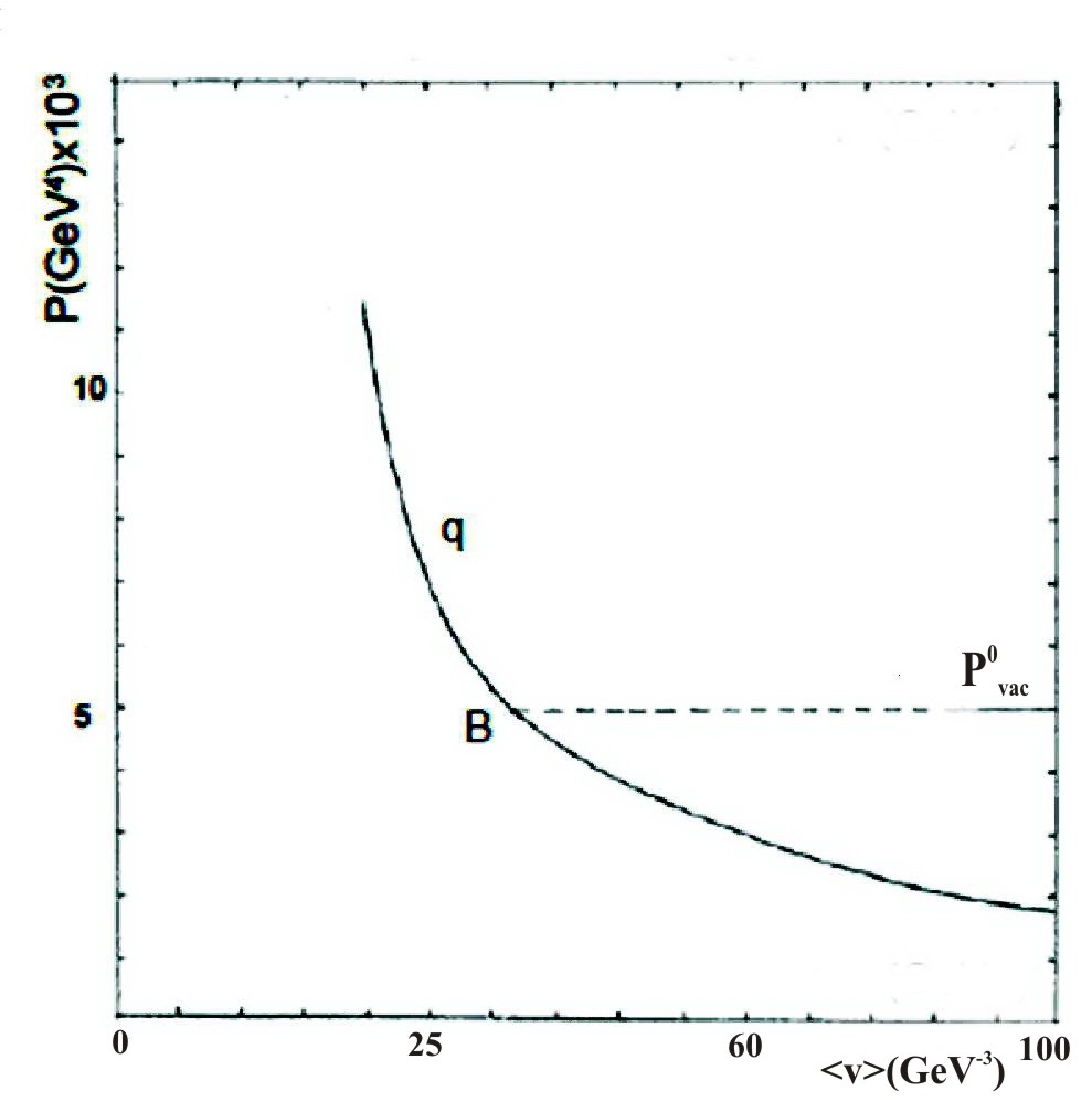}
\caption{The pressure of the HPh non-perturbation QCD vacuum
  condensate (horizontal segment $P^0_{vac}$) vs the pressure of the  
  degenerate (''cold'') perfect gas of $(u,d,s)$ current quarks 
  (curve $q$). As the particle number density approaches the critical value 
  (the neutron spacing 
  becomes compact just as the quark specific volume is
  $\langle v\rangle\,\simeq$ 100 GeV$^{-3}$), the occurrence of a gross gap 
  between the HPh- and SHPh-phase pressures is unambiguously pronounced - at 
  $\langle v\rangle\,\simeq$ 100 GeV$^{-3}$ the 
  former is about three times as large as the latter one.  }
\end{figure}

\newpage

2.\quad{\bf\boldmath $Soft\,\,scenario$: \quad No stepwise HPh 
$\longleftrightarrow$ SHPh transition}\\
 
This scenario implies that an intermediate state is to be passed after the 
neutrons ''get 
in touch with each other''. Namely, first, the neutrons disintegrate into the 
deconfined massive dynamical quarks (valons) \cite{Sh,F,F-Ch,Cley,AKM} and, 
then, both the valon masses 
and vacuum condensate pressure decrease similarly along with the medium density 
increase \cite{R_2009}; finally, the valons turn into the current quarks and, 
thus, the vacuum condensate vanishes, $P_{vac}\,=\,-\varepsilon_{vac}\,\to\,0$, 
more or less gradually. 

A reasonable approach was suggested \cite{R_2009} which would describe the 
degenerate valonic gas at the particle energy densities 
$\varepsilon\,\geq\,|\varepsilon^0_{vac}|$ \footnote{If a steady state of this 
type were ever accessible - we shall argue below that, actually, it is not the 
case.}. It is based on the EoS
for a perfect gas with effective particle mass varying along with the medium 
density variation:

\begin{equation} 
\varepsilon\,=
\frac{6 N_f}{2\pi^2}\,\int_0^{p_F} dp\,p^2 \sqrt{p^2 + m^2 (\varepsilon)},
\end{equation}

where $N_f$ =3 is the number of flavors allowed for and the Fermi 
momentum $p_F\,=\,(\frac{\pi^2}{N_f \langle v\rangle})^{1/3}$. Let us
parameterize the valon mass as

\begin{equation}
m_{u,d}\,\simeq\,m_0\,\exp[-a\,(\varepsilon/|\varepsilon^0_{vac}|\,-\,1)],
\end{equation}  

and, correspondingly,

\begin{equation}
\varepsilon_{vac}\,\equiv\,-P_{vac}
\simeq\,\varepsilon^0_{vac}\,\exp[-a\,(\varepsilon/|\varepsilon^0_{vac}|\,-\,1)]
\end{equation} 

where $m_0\,\simeq\,\frac{1}{3}m_n\,\simeq$ 330 MeV \footnote{Actually, the 
$\sim$150-MeV mass-difference between $(u,d)$- and $s$-valons was allowed for, 
but no significant 
correction was shown to come therefrom \cite{R_2009}.} and $a$ 
is a free parameter, which describes the rate of QCD vacuum 
condensate destruction. The numerical solutions of eq.(1), supplemented 
with eq's.(2,3) and the thermodynamic relation 
$P\,=\,-\partial(\langle v\rangle \varepsilon)/\partial\langle v\rangle$, 
are presented in Fig.2. Note, that only values $a\,\geq$\,1 are 
physically reasonable 
because the HPh vacuum condensate should be significantly affected as the 
particle energy density approaches the absolute value of the 
condensate strength itself (or even earlier). The curves 
2-4, which refer to $a\,<$\,1, are depicted for an illustration only. It is 
evident that hard scenario comes back in the limit $a\,\to\,\infty$.

\newpage

 \begin{figure}[h]
\includegraphics[width=10cm]{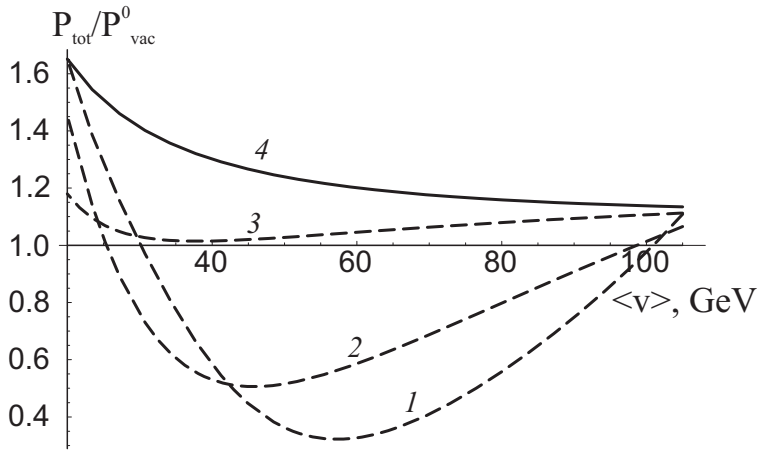}
\caption{{Formal solutions for the soft-scenario EoS at 
different rates of phase 
transition (HPh $\longrightarrow$ SHPh) in the ''$cold$'' (degenerate) nuclear
matter. The curves (1 - 4) refer to the different regimes of the vacuum 
condensate destruction as well as of the valon mass decrease along with growing 
up the particle (valon) 
energy density $\varepsilon$ (at $a$ = 1, 0.5, 0.1, 0.01, respectively). 
One can see that a steady state regime of ''$cold$'' nuclear matter compression 
($dP_{tot}/d\langle v\rangle\,<$ 0) is quite 
improbable:
it asks for unreasonably robust vacuum condensate, which would keep itself
almost unchanged until the particle energy density becomes, at least, about one 
order higher than the module of HPh-condensate energy density 
$\varepsilon^0_{vac}$ ($a\,\leq\,0.1$). 
Within the defensible framework, which implies that this condensate is 
subjected to significant destruction as both energy densities become of the 
same order (the downmost curve 1, $a = 1$) or even earlier, the inequality 
$dP_{tot}/d\langle v\rangle\,>$ 0 would hold inevitably for a certain stage to 
be passed under the gravitational compression. This fact
signals, undoubtedly, of instability. In other words, 
the nuclear substance can no longer remain ''cold'' (degenerate).}}
\end{figure}

\vspace*{0.5cm}     

Thus, basing on the set of solutions shown in Fig.2, we come to the principally 
significant conclusion that {\it no way 
is conceivable for ''cold'' HPh $\longrightarrow$ SHPh  transition}
\footnote{Observed EoS softening towards the center of large mass NS
\cite{soft} could be considered as a certain phenomenological manifestation in 
favor of this statement, although the authors themselves suggest a different 
reasoning for this fact.}.  

\newpage

\subsection{Blowing up NS \, {\it versus} \, BH formation}

Nevertheless, one can expect that a certain transient ''quasi-steady'' state of 
NS (instead of the immediate NS rupturing) could be still maintained for some 
range of large $M_{NS}$, despite of enormous thermal disbalance between some
small hot central domain and the ''rest'' of the star.  

Is this pattern favorable for leaving NS a loophole to evolve towards the final
BH configuration due to fast gravitational compression, which could 
enclose the horizon regardless of blowing up made by the divergent heat flow? 
Below, we try to put forward some meaningful 
arguments that the most reasonable answer must be negative.  

As being emerged at the star center, the SHPh domain starts swelling until a 
transient balance is established between further heating due to gravitational 
compression and a pretty slow heat outward transport \footnote{Note again that
neutrinos get essentially stuck at the relevant densities of nuclear matter and, 
therefore, this transport is an extremely slow process (tens of hours {\it vs}
the typical hydrodynamic-time scale, which is, probably, of some milliseconds).}. 
The inner (SHPh) domain is a kind of very hot subhadronic matter, which is 
called below, for brevity, QGP \footnote{In the 
present context, it is referred to be a nearly perfect gas, which consists of 
the unremovable ''primordial'' quarks (the net 
baryon-over-antibaryon surplus) as well as of the multiply produced qluons and
$q\bar q$-pairs, baryonic chemical potential $\mu_B$ thus tending to zero.
Actually, the reasoning we put forward keeps valid for any microscopic
structure, which mimics macroscopically the perfect gas thermodynamic 
behavior.}. 
If the temperature of this plasma is $T$, then the obvious energy-conservation
equation reads:

\begin{equation}
-\,AG \frac{M_{NS}^2}{R^2} dR\,\simeq\,
4\pi \sigma_{QGP} T^4\,(1\,+\,
\frac{|\varepsilon^0_{vac}|\,-\,\varepsilon_n}{\sigma_{QGP} T^4}) r^2 dr,
\end{equation}

where on the left-hand side stands the work made by the gravitational field
($M_{NS}$ and $R$ are the NS mass and its radius, respectively, and
the value of coefficient $A$ is 
confined in between of its non-relativistic and ultra-relativistic limits, 
$\frac{6}{7}\,\leq\,A\,\leq\,\frac{3}{2}$ \footnote{Below, we put $A\,=\,1$, 
since, in fact, the ultra-relativistic limit is rather inaccessible for the 
HPh-medium.}), while on the right-hand side stands the energy increase within 
the domain of radius $r$ occupied by 
SHPh (QGP),   

$$\sigma_{QGP}\,=\,\frac{\pi^2}{30}(2\,\times\,8\,+
2\,\times\,3\,\times\,2\,\times\,3\,\times\,\frac{7}{8})$$

being the 3-flavor QGP weight factor (8 gluons of spin 1 and (3 + $\bar 3$) 
colored quarks of spin 1/2).

Nearly outside of the phase transition ''boundary'' 
(which is, actually, not a boundary but rather an extended spherical layer),
the energy density of HPh substance made of closely packed neutrons approaches 
the value $\varepsilon_n\,\simeq\,|\varepsilon^0_{vac}|$, what 
refers to $a\,\simeq$ 1, in eq.s (2,3). This is just about the total 
density because the QCD vacuum condensate tends to 
zero. On the other hand, on the layer inside, it equals to the energy density of 
high-temperature QGP.  Thus,    
the hydrodynamic (fast process) balance asks strongly for the leveling of two 
these energy densities: 

\begin{equation}
|\varepsilon^0_{vac}|\,\simeq\,\sigma_{QGP} T^4,
\end{equation}

wherefrom one obtains $T\,\simeq$ 130 MeV \footnote{This is, at least, one 
order higher
than the typical temperatures for the supernovae explosions.}, and we see that 
this balance could be only
maintained at the price of an enormous thermal disbalance (remind that the NS
medium temperature outside of the phase transition ''boundary'' is of a few MeV). 
It is worth emphasizing that this estimate is quite compatible with the 
$\mu_B$ = 0 lattice MC simulation result \cite{Karsch} for HPh $\to$ SHPh
transition, which was found to be a crossover lasting over the temperature
interval $140 \mbox{MeV}\,\leq\,T\,\leq\,200$MeV. 

Basing on eq.(5), one can reasonably assume that the second term in the brackets 
on the right-hand side of eq.(4) is small as compared to unity. If so, then the
transient ''quasi-steady'' mode of a high-mass NS nuclear medium is composed as 
follows:  

\begin{equation}
G \frac{M_{NS}^2}{R}\,=\,\simeq\,
\frac{4\pi}{3} \sigma_{QGP} T^4 r^3\,+\,C,
\end{equation}

where $C$ is defined by $\overline{M_{NS}}$ - the value of mass upper limit 
for the {\it really} 
stable (''cold'', i.e., $r$ = 0) NSs: $C\,\simeq\,(0.5\,\div\,1)\,M_{\bigodot}$
for $\overline{M_{NS}}\,\simeq\,(1.5\,\div\,2.5) M_{\bigodot}$ 
and $R\,\simeq\,(8\,\div\,10)$ km, respectively. Of course, the correlation (6) 
between $M_{NS}$ and $r$ can be defensible at $r\,\ll\,R$
only. In this case, a rather ''peaceful'' 
evolution of NS is not ruled out; it gives rise to the production of some
cannonballs and/or successive GRBs, which become, however, the more destructive 
the larger is $M_{NS}$, thus resulting in diminishing the NS mass until it 
approaches the upper limit $\overline{M_{NS}}$ of NS stability. Otherwise (at 
still higher NS masses, when eq.(6) would predict $r$ and $R$ of the same order), 
no transient hydrodynamic balance is conceivable at all - the development of 
powerful shock waves seems inevitable which should forward NS towards the 
catastrophic self-destruction \footnote{From the more general point of view, all
that is nothing else than different ways of symmetry (in this case - the chiral
one) breaking along with the medium cooling: the no-order-parameter SHPh turns 
into the HPh, which shows up clearly an order parameter - it can be chosen to be
the inverse radius of color confinement.}.


At the same time, the elementary condition for horizon first appearance within 
the body of a compact star reads: $\frac{2GM_g}{R_g}$ = 1, or

\begin{equation}
R_g\,\simeq\,\lbrack\frac{3}{8\pi 
G \langle\varepsilon_g \rangle }\rbrack^{1/2,}
\end{equation}

where $R_g$ and $\langle\varepsilon_g \rangle$ are the BH radius and its mean 
energy density, respectively. For getting the lower estimate of $R_g$, one 
has to take into account that 
$\langle\varepsilon_g \rangle\,\leq\,|\varepsilon^0_{vac}|$, since, otherwise, 
the phase transition instability followed by the aforementioned destructive 
cataclysms is expected to activate before. Thus, one obtains

\centerline{$R_g\,\geq\,12$ km or $M_g\,\geq\,4\,M_{\bigodot}$}  
\vspace*{3mm}

We see that the admissible maximal NS and minimal BH radii are rather 
compatible, while the corresponding masses are separated by a significant gap.
What is in between? If the NS mass were imagined to access $4\,M_{\bigodot}$, 
then eq.(6) tells 
immediately that $r\,\simeq\,R$, what is, actually, meaningless.  Therefore, 
the only interpretation, which remains within eyeshot, is that the star can not
''jump over'' the gap without being ruined in full.

As for the stars of lower $\langle\varepsilon_g \rangle$ 
(larger $M_g$ and $R_g$,
both being $\sim\,\langle\varepsilon_g \rangle^{-1/2}$), the question remains 
open, since, in this case, a more detailed description of star dynamics should
be involved. What can be said, is that shutting to BH seems to be, actually, 
an event of even lower probability than 
it might be thought of using the above consideration. The matter is that the 
horizon appearance is linked to the $global$ features of the star nuclear 
medium (the value of $M/R$ and $averaged$ energy density 
$\langle\varepsilon \rangle$), while the HPh $\longrightarrow$ SHPh  
transition instability is linked directly to the $local$ value of 
$\varepsilon$, which is, undoubtedly, $r$-dependent and increases towards the 
star center. The same argument gives, all the more, ''obvious preference'' 
to the proactive development of HPh $\longrightarrow$ SHPh instability in case 
of some density fluctuations within the star body. 
Thus, this instability is anyway expected to start developing at lower values 
of $\langle\varepsilon \rangle$.  

It is worth also 
mentioning, in this connection, that a number of additional factors - possible 
star non-sphericity, star rotation and, especially, binary-star configuration -
should, obviously, result in diminishing the margin of NS 
stability, thus making the 
above arguments against a BH-horizon appearance even more defensible.

\section{Conclusion}
The QCD-induced mechanism of additional NS instability is discussed. The NSs of 
highest masses are proven to be in face of instability associated with 
QCD-vacuum transformation under HPh $\longrightarrow$ SHPh transition, which 
could manifest itself, in particular, through the softening of EoS towards the 
star center. This instability seems to develop before a BH horizon appears 
within the star body, what makes rather improbable the very 
accessibility of a BH configuration at the end of collapsing star evolution.\\

Since the SHPh-phase temperature is, at least, one order higher than that of 
the supernovae explosions, the expected energy release could be several 
(seemingly, up to 2-3) order higher. That is why it is difficult to resist the 
temptation of 
linking the QCD-induced instability under discussion and the poorly understood
data on very distant (''young'') GRB's of highest energy, 
like GRB 090423 \cite{krimm}, GRB 080916C \cite{abdo}, GRB 080319B  
(''naked eye'') \cite{bloom}, etc.      


\end{document}